\documentclass[twocolumn,prl,amssymb,epsfig,aps,lettersize]{revtex4}
\usepackage{epsfig}
\def\ncalcs{226\,}

\begin{document}
\textheight 227mm 

\title{\Large {Structure maps for hcp metals from first principles calculations}}
\author{ Ohad Levy$^{1,2}$, Gus L. W. Hart$^{3}$, and Stefano Curtarolo$^{1,4,\star}$}
\affiliation{
  $^1$Department of Mechanical Engineering and Materials Science and Department of Physics, Duke
  University Durham, NC 27708 \\
  $^2$Department of Physics, NRCN, P.O.Box 9001, Beer-Sheva, Israel\\
  $^3$Department of Physics and Astronomy, Brigham Young University, Provo UT 84602\\
  $^4$Department of Materials and Interfaces, Weizmann Institute of Science, Rehovot 76100 Israel \\
  $^\star${corresponding author, e-mail: stefano@duke.edu}
} \date{\today}
\begin{abstract}
  The ability to predict the existence and crystal type of ordered structures of materials
  from their components is a major challenge of current materials research.
  Empirical methods use experimental data to construct structure maps
  and make predictions based on clustering of simple physical parameters. Their
  usefulness depends on the availability of reliable data over the entire
  parameter space. Recent development of high throughput methods opens the
  possibility to enhance these empirical structure maps by {\it ab initio}
  calculations in regions of the parameter space where the experimental evidence is lacking
  or not well characterized. In this paper we construct enhanced maps for the binary alloys
  of hcp metals, where the experimental data leaves large regions of poorly
  characterized systems believed to be phase-separating. In these enhanced maps, the
  clusters of non-compound forming systems are much smaller than indicated by the empirical results alone.
\end{abstract}
\maketitle

\section{Introduction}
Predicting the stable structures of alloys from their constituents is a major challenge of current
materials research \cite{Maddox,Woodley}. The traditional approach to this problem is to extract
trends from systems for which experimental data is available and apply them to uncharacterized
systems \cite{villars:factors}. These rules usually depend on simple parameters, e.g.\ atomic
number, atomic radius, electronegativity, ionization energy, melting temperature or enthalpy,
etc. Several well-known methods include the Hume-Rothery rules \cite{hume_rothery}, the Miedema
formation enthalpy method \cite{Miedema}, the Zunger pseudo-potential radii maps \cite{zunger:1980}
and Pettifor maps \cite{pettifor:1984,pettifor:1986}.

These methods produce results largely consistent with existing experimental data and have helped
direct a few successful searches for unobserved compounds \cite{pettifor:2003}. However, they offer
a limited response to the challenge of identifying new compounds because they rely on the existence
of consistent and reliable information for systems spanning most of the relevant parameter space. In
many cases, reliable information is missing in a large portion of the parameter space---less than
50\% of the binary systems have been satisfactorily characterized \cite{Villarsetal_JAC01}. This
leaves considerable gaps in the empirical structure maps and reduces their predictive
ability.

High-throughput (HT) calculations of material properties based on density functional theory (DFT)
have been developed for theoretically guided material discovery and design
\cite{Johann02,Stucke03,curtarolo:prl_2003_datamining,monster,Fischer06,Lewis2007355,ozolins:135501}. 
In this approach the phase stability landscape of alloys is explored by calculating the formation
enthalpies of a large number of possible structures, identifying the minima at various component
concentrations. A minimum-free-energy convex hull is constructed from these minimum energy structures.
These calculations give insights into trends in alloy properties and indicate possible existence of
hitherto unobserved compounds. Recent advancements in this method, by efficiently covering extensive lists
of candidate structure types \cite{JACS_Rh_2010,curtarolo:art51,curtarolo:art49,curtarolo:art55},
make it possible to complement sparse experimental data
with {\it ab initio} total energy assessments. This development was envisioned by Pettifor
a few years ago \cite{pettifor:2003} and here we present its first realization for binary metallic
systems.

\begin{figure}
   \includegraphics[width=\linewidth]{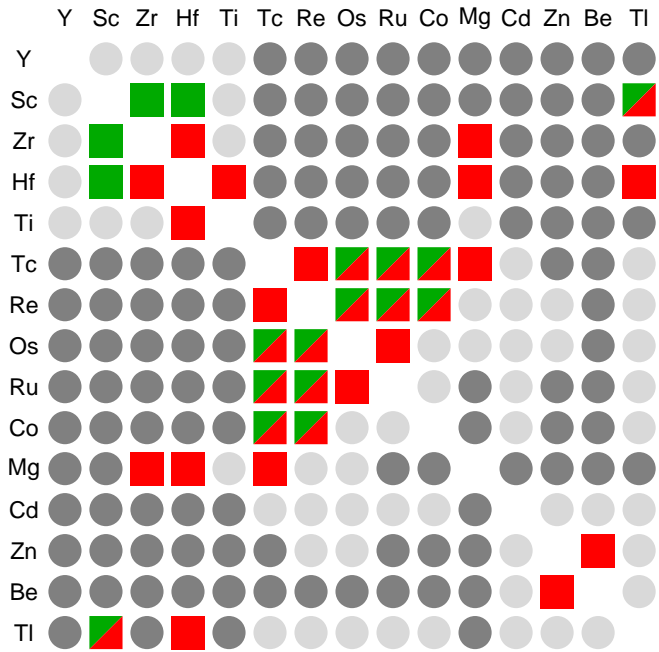}
  \caption{\small
    (Color online)
    Compound-formation or phase-separation in binary systems of the hcp metals,
    ordered by the Pettifor $\chi$-scale.  Circles denote agreement of the empirical data and HT
    calculations on a database of $\sim$\ncalcs\, structures, for phase separating (light gray) and
    compound-forming (dark gray) systems. Squares denote HT prediction of stable ordered structures
    in systems reported as phase-separating (green, red or mixed for hcp, non-hcp or both types of
    predicted structures, respectively).}
  \label{fig1}
\end{figure}

We apply the HT approach to a comprehensive screening of the 105 intermetallic binary systems of the
hcp metals. These systems have been generally less extensively studied than the cubic (fcc and bcc)
metals, both experimentally and theoretically. The experimental data is particularly sparse for a
large group (46) of binary intermetallics believed to be phase separating (see Figure
\ref{fig1}). {\it Ab initio} calculations on these systems are scarce compared to bcc and fcc
systems. 
Cluster expansion \cite{sanchez_defontaine:prb_1982_ce,sanchez_ducastelle_gratias:psca_1984_ce}
studies of hcp systems \cite{McCormack,PhysRevB.5.1591,rmccormackMRS,PhysRevB.48.748} are also relatively rare
compared to cubic systems, probably due to the inherent additional computational difficulties in
dealing with a lattice having  a basis of symmetrically identical atoms 
(a general cluster expansion code for arbitrary parent lattices was not available until 2002 \cite{atat1}).

\section{Methods}
Computations of low temperature stability of the hcp metallic binary systems were carried out using
the HT framework {\small AFLOW} \cite{monster,aflow}
and {\it ab initio} calculations of the energies with the {\small VASP} software \cite{kresse_vasp} with
projector augmented waves (PAW) pseudopotentials \cite{paw} and the exchange-correlation
functionals parameterized by Perdew, Burke and Ernzerhof for the generalized gradient approximation
\cite{PBE}. The energies were calculated at zero temperature and pressure, with spin polarization
and without zero-point motion or lattice vibrations.
All crystal structures were fully relaxed (cell volume and shape and the basis atom
coordinates inside the cell). Numerical convergence to about 1 meV/atom was ensured by a high energy
cutoff (30\% higher than the maximum cutoff of both potentials) and dense 6000 {\bf k}-point Monkhorst-Pack
meshes \cite{monkhorst}.

For each system, we calculated the energies of the reported crystal 
structures \cite{Pauling,Massalski} and approximately \ncalcs\,\, additional structures in the AFLOW database \cite{aflow}.
These include the 176 configurations described in \cite{monster}, all the
symmetrically-distinct hcp-, bcc-, fcc-based superstructures
\cite{gus_enum,multienum} with up to four atoms per cell, and the
prototypes A5, A6, A7, A8, A9, A11, B20, C36, D5$_{19}$,
Al$_2$Zr$_4$, Al$_3$Zr$_2$,
BiHf$_2^{\star}$ \cite{curtarolo:art51},
CdTi, CuPt$_7$, Cu$_3$Ti$_2$,
Ga$_2$Hf, Ga$_4$Ni, Ga$_3$Pt$_5$, Ga$_4$Ti$_5$,
Hf$_3$Sc$^{\star}$ \cite{curtarolo:art51}, Hf$_5$Sc$^{\star}$ \cite{curtarolo:art51},
Hf$_5$Pb$^{\star}$ \cite{curtarolo:art51}, HfPd$_5^{\star}$ \cite{curtarolo:art51},
Hf$_2$Tl$^{\star}$ \cite{curtarolo:art51},
Hg$_2$Pt,
ITl, InTh,
LiB-MS1/2 \cite{kolmogorov:binary_LiB_2006,curtarolo:art26},
MoTi$^{\star}$ \cite{monster}, Mo$_3$Ti$^{\star}$ \cite{monster}
NbNi$_8$, NiTi$_2$, SeTl and V$_4$Zn$_5$.
In Ref. \cite{monster}, it
was shown that the probability of reproducing the correct
ground-state, if well defined and not ambiguous, is 
$\eta_c^\star \sim 96.7\%$ ({\it ``reliability of the method''}, Eq. (3)). There is no
guarantee that the true ground states of a system will be found among
the common experimentally observed structures or among small-unit-cell
derivative structures. However, even if it is impossible to
rule out the existence of an unexpected ground-state, this protocol
(searching many enumerated derivative structures and exhaustively
exploring experimentally reported structures) is expected to give a
reasonable balance between high-throughput speed and scientific
accuracy to determine miscibility (or lack thereof) in these alloys.

\section{Results and Discussion}
Of the 105 hcp-hcp metallic binary systems, 46 were previously reported as phase-separating
\cite{Pauling,Massalski}. Our calculations show that 18 of those are actually compound-forming at
low-temperatures. Most of the predicted compounds are \emph{not} hcp-derived structures. These
results are summarized in Figure \ref{fig1}, where all the hcp metals are ordered according to
Pettifor's chemical scale $\chi$ \cite{pettifor:1984,pettifor:1986}. The symbols in each
intersecting cell indicate the phase-separating or compound-forming nature of the binary alloys.
Circles denote cases in which the HT calculation results match the experimental data for phase
separating (light gray) and compound-forming (dark gray) systems. Alloys where discrepancies are
found, i.e. alloys reported as phase-separating but predicted to have stable ordered structures, are
denoted by squares, green - if the predicted structures are hcp-derived, or red - if they are not
hcp-based. Squares with both colors indicate a prediction of both types of ground-states in the same
system. No reverse discrepancies were found, i.e.\ no cases were found where systems reported to
be compound forming were predicted to be phase separating at low temperatures.

Most of the 46 reported phase-separating systems are grouped in this map into three clusters. 10
systems in the upper left corner, 6 in the lower right corner and 23 in a large cluster in the
center. The last metal in the map, thallium, has 10 phase separating systems, 7 of which stand out
of these three clusters. The {\it ab initio} results predict 4 compound-forming systems grouped in
the center of the upper left cluster, which is reduced to just 6 systems. The central cluster is
also reduced to just 12 systems. In particular, Mg-Zr and Hf-Mg, which form an extended branch of
this cluster are predicted to be compound forming. The new smaller cluster is thus better localized
in the chemical scale space.
\begin{figure*}[t]
  \includegraphics[width=\linewidth]{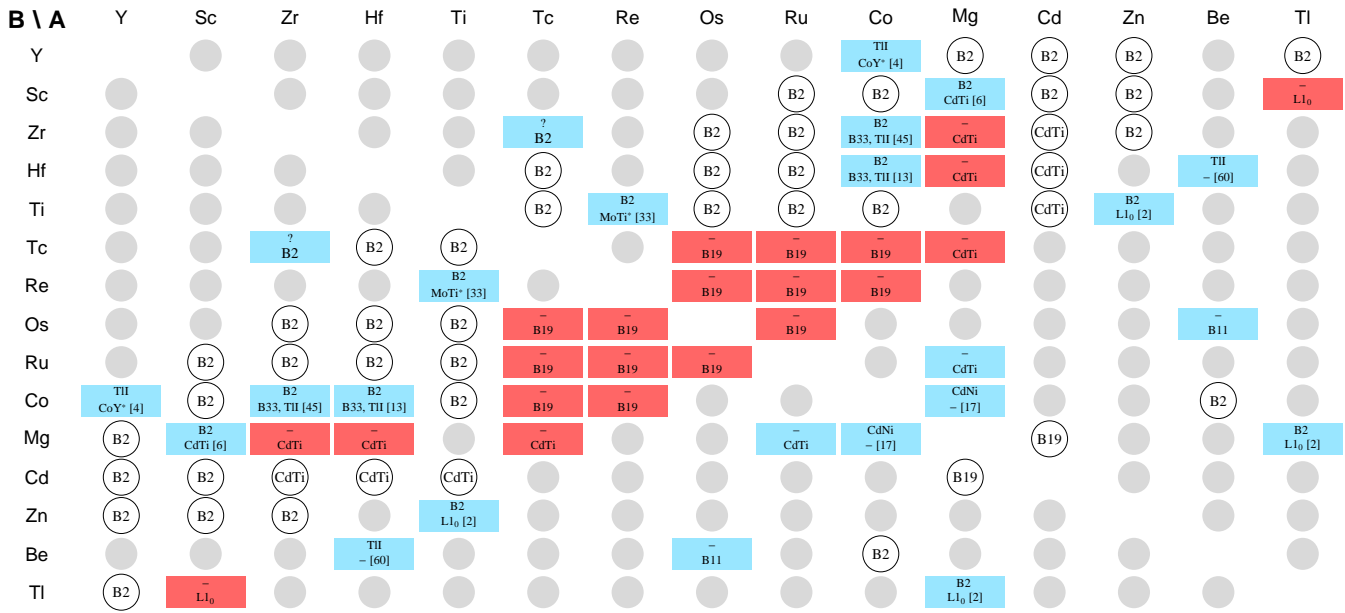}
  \caption{\small
    (Color online)
    The AB structure map for hcp alloys, from empirical data (upper entry in each cell) and HT calculations (lower entry). Circles
    with a single entry indicate agreement between experiment and calculation.
    Predicted ordered structures in reported phase-separating systems are highlighted in red. Discrepancies between observed and
    calculated structures are highlighted in blue. Suspected compounds with unknown prototype are
    denoted by a question mark. New prototypes are marked by $\star$
    and described in Table \ref{table_protos}. The energy difference between reported and calculated structures or
    between the reported structure (unstable in the calculation) and a two-phase tie-line is indicated in
    square parentheses.}
  \label{fig2}
\end{figure*}
\begin{figure*}[t]
  \includegraphics[width=\linewidth]{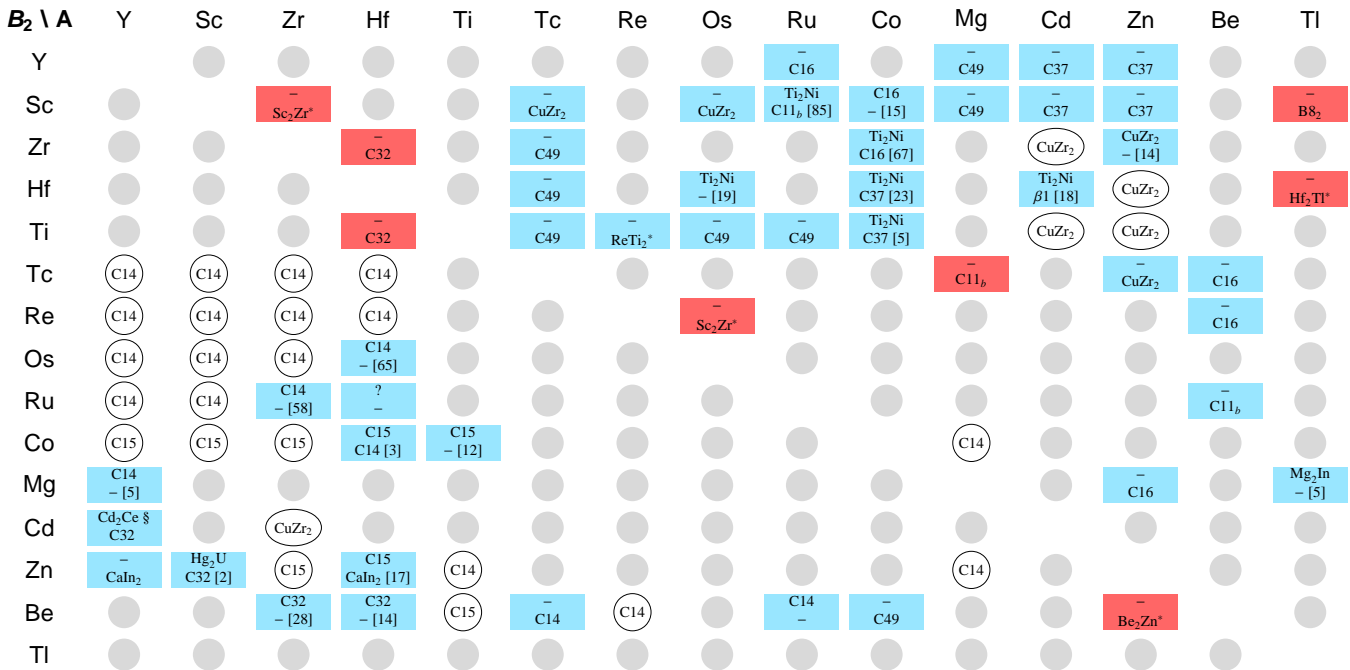}
  \caption{\small
    (Color online)
    The AB$_2$ structure map for hcp alloys, from empirical data and HT calculations.
    Notations are as in Figure \ref{fig2}. $\S$ indicates that the reported prototype relaxed to the
    predicted one in the calculation.}
  \label{fig3}
\end{figure*}
\begin{figure*}[t]
  \includegraphics[width=\linewidth]{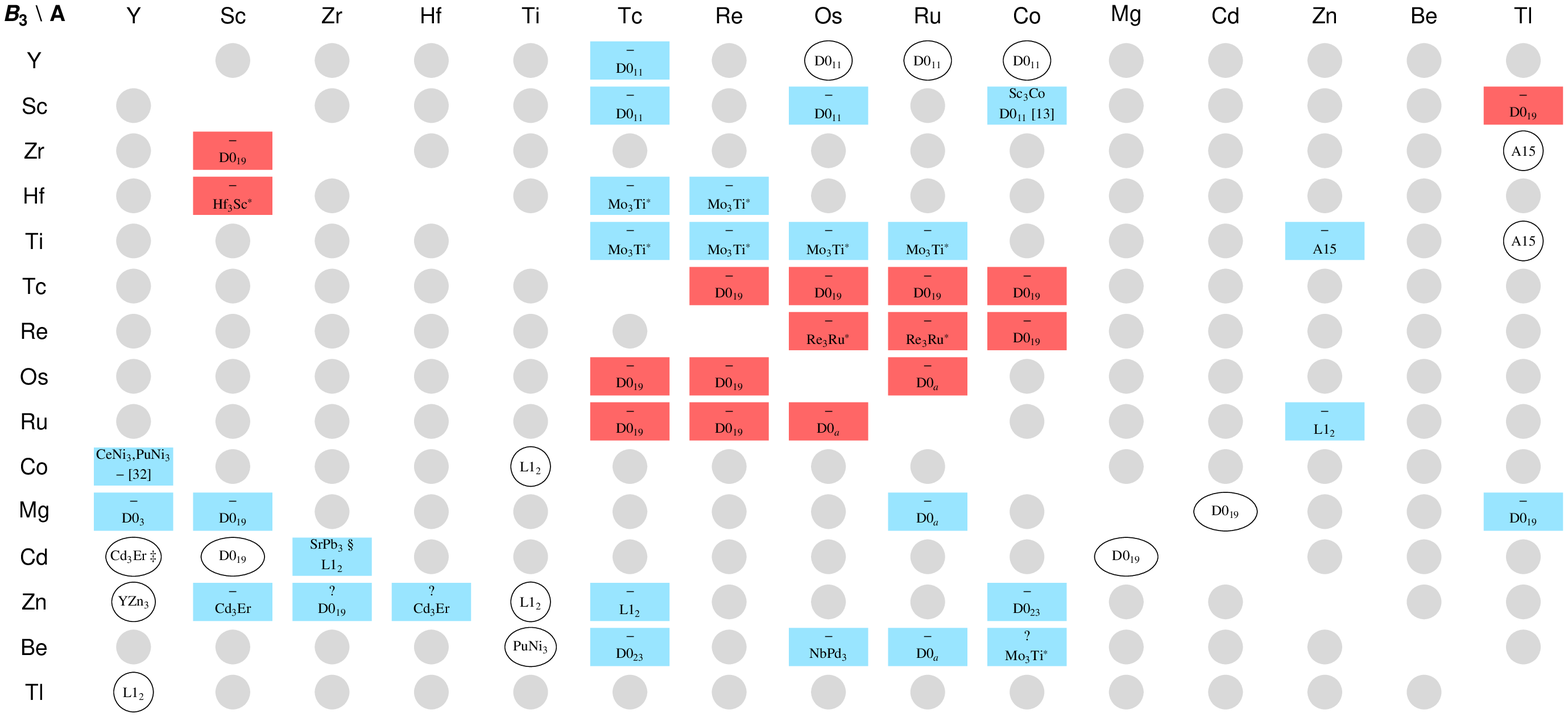}
  \caption{\small
    (Color online)
    The AB3 structure map for hcp alloys, from empirical data and HT calculations.
    Notations are as in Figure \ref{fig3}. $\ddagger$ In the case of, YCd$_3$, the high throughput
    results find that the Cd$_3$Er structure and D$0_{19}$ have the essentially same energy.}
  \label{fig4}
\end{figure*}
Alternative Mendeleev number assignments are sometimes suggested (e.g., see
Ref.~\onlinecite{Villarsetal_JAC01}). One such suggestion places Mg and Be adjacent to each other
before Sc and Y. The map in Fig. \ref{fig1} indicates that the ordering according to the Pettifor
$\chi$ scale locates them more appropriately at the lower part of the scale.  A few of these
alternative arrangements also place Sc before Y, whereas our calculations indicate that the $\chi$
scale is again more appropriate, since it groups together the predicted compound-forming systems in
the upper left cluster.  Two isolated phase-separating systems, Sc-Tl and Hf-Tl, are predicted to be
compound-forming, so it now seems that Tl should be more appropriately placed between Cd and Zn. The
calculation results thus complement the empirical Pettifor map and group all the phase-separating
systems into three small well-defined clusters.

Figures \ref{fig2}--\ref{fig4} show the resultant structure maps $(\chi_A,\chi_B)$ for the binary
AB, AB$_2$ and AB$_3$ compounds.  The ordered structures are represented by their {\it
 Strukturbericht} designation or prototype. A circle indicates agreement of the empirical data and
the HT calculations. Cases of disagreement are indicated by rectangles. The upper and lower entries
in each cell give the empirical data and the HT result, respectively. In cases where two different
structures are reported in the literature or found degenerate in the calculations, both are
indicated as upper or lower entries. A dash (-) indicates the absence of a stable
structure. Unidentified suspected structures are denoted by a question mark (?). Some of the
predicted phases (marked by an asterisk *) have structures for which no prototype is known and no 
{\it Strukturbericht} designation have been given. These new prototypes are described in Table
\ref{table_protos}.

Systems reported as phase-separating but predicted by our HT results to be compound forming are
highlighted in red. In systems reported to be compound-forming, discrepancies between observed and
calculated structures are highlighted in blue. In these cases, the energy difference between the
structures or between the reported structure (unstable in the calculation) and the two-phase
tie-line is indicated in square parentheses.

Most of the predicted compounds in the AB map are concentrated in a large cluster at the center, in
the largest gap of the empirical map. Seven of them form a new cluster of B19 structures. Three
Mg compounds of the CdTi prototype neighbor the reported group of Cd compounds with the same
structure. The main feature of the empirical AB map is a large cluster of B2 compounds. 
The {\it ab initio} calculations largely preserve this cluster and the separation between it and the adjacent
group of CdTi compounds. Discrepancies between reported and predicted structures appear mostly at
the margins of these clusters. For example, the MgSc compound, reported as B2, is predicted to
switch to the adjacent CdTi cluster at low temperatures. An unidentified TcZr compound (denoted by a
question mark) is predicted to belong to the B2 group and an unreported MgRu compound is predicted
to belong to the CdTi cluster. Most of the energy differences between reported and predicted
structures are small. Relatively large differences are found for CoZr and CoHf, where the reported
B2 structure is predicted to be replaced by degenerate B33/TlI structures, and for ReTi, where an
unreported prototype, MoTi \cite{monster}, is predicted to replace B2 at low temperatures.

The structure map of the AB$_2$ compounds is shown in Figure \ref{fig3}. In this map
there are eight predicted structures in reported phase-separating systems. Five of them appear in the
Sc-Zr, Hf-Zr, Hf-Ti, Hf-Tl and Be-Zn systems for which there are none in the AB map (Figure \ref{fig2}).
The main features of the empirical map are maintained, which are a large cluster of C14 compounds and
a smaller C15 group bordering it in the lower left part and a small CuZr$_2$ group on the upper left. 
Discrepancies between reported and calculated structures appear at the margins of these clusters. 
In addition, small groups of C49 and C37 compounds are predicted in the upper part.

Figure \ref{fig4} shows the structure map for the AB$_3$ compounds. Predicted structures appear in
the Hf-Sc and Re-Tc systems, for which there are none in the AB and AB$_2$ maps (Figures \ref{fig2}
and \ref{fig3}). Thus, all the systems reported as phase-separating but predicted to be
compound-forming have ordered structures in at least one of these three maps. In addition, a large
cluster of 13 structures is predicted in the non compound-forming gap in the center of the
map. Eight of these are D0$_{19}$ structures and two have a closely related new hcp-prototype
Re$_3$Ru (and OsRe$_3$). There are no major features in the AB$_3$ empirical map except a small group
of D0$_{11}$ Y-compounds in the upper part. This group is reproduced and extended in the
calculations. The reported prototype Cd$_3$Er, of Cd$_3$Y, is found to be degenerate with the D0$_{19}$ structure.

Also predicted is a cluster of Hf and Ti-compounds of the unreported prototype Mo$_3$Ti
\cite{monster}. This concentration of distorted bcc structures (Mo$_3$Ti is an oI8 distortion of a
bcc superlattice, Table \ref{table_protos}) in six Hf and Ti binary systems may reflect their strong
affinity to bcc environments. At zero temperature, bcc-Ti is dynamically unstable having an
imaginary frequency phonon branch at the N point \cite{PetryPRB1991_Ti,Nishitani200177}, but it
stabilizes at 882$^\circ$C, where Ti undergoes a hcp-bcc transition \cite{Pauling}. Hence, an
introduction of an alloying agent that stabilizes this phonon branch, can push the Ti alloy into a
bcc-type environment even at low temperature. Hf also undergoes an hcp-bcc transition at a much
higher temperature of 1743$^\circ$C, \cite{Pauling}. It is known, however, that alloying elements
such as Os, Re and Mo reduce this transition temperature and increase the stability domain of the
bcc-phase \cite{Massalski}. The calculations show
that a few alloying agents may be able to drive this extension down to very low temperatures.

\begin{widetext}
  
  \begin{table}[htb]
    \caption{
      Geometry of new prototypes (marked by ``$^{\star}$'' in Figures \ref{fig2},\ref{fig3},\ref{fig4}). 
      Atomic positions and unit cell parameters are fully relaxed.}
    \label{table_protos}
    \scriptsize
    {
      
      \hspace{-8mm}
      \begin{tabular}{||c|c|c|c|c||}\hline\hline
        System                              &  CoY$^{\star}$        &  Sc$_2$Zr$^{\star}$      & MoTi$^{\star}$\cite{monster}          & Mo$_3$Ti$^{\star}$\cite{monster} \\ \hline
        Lattice                             &Rhombohedral           &   Monoclinic             &  Orthorhombic          & Orthorhombic                                            \\ \hline
        Space Group                         & R$\bar{3}$m \#166     &   C2/m \#12              &  Imma \#74             &Immm \#71                                                \\ \hline
        Pearson symbol                      & hR4                   &    mS12                  &      oI8               &          oI8                                            \\ \hline
        Prim. vect.                         &  (hexagonal axes)     &  (SG option 1)           &  (SG option 2)         &                                                         \\
        $a,b,c$ (\AA)                       &3.896, 3.896, 20.609   &11.800, 3.271, 7.623      &4.479,3.182,9.046       &   4.444, 3.173, 8.971                                   \\
        $\alpha,\beta,\gamma$ (deg)         &   90, 90, 120         & 90, 76.744, 90           &90, 90, 90              &90, 90, 90                                               \\ \hline
        Wyckoff                             & Co1 0,0,0.3326 (6c)   &Sc1 0.1914,0,-0.1370 (4i) &Mo1 0,1/4,0.3847 (4e)   &Mo1 0,0,0.2440 (4i)                                      \\
        positions                           & Y1 0,0,0.0869 (6c)    &Sc2 0.1334,0,0.4740 (4i)  &Ti1 0,3/4,-0.1343 (4e)  &Mo2 0,1/2,0 (2d)                                         \\
        \cite{bilbao,tables_crystallography}&                       &Zr1 -0.4739,0,0.1931 (4i) &                        & Ti1 1/2,0,0 (2b)                                        \\  \hline
        {\small AFLOW} label \cite{aflow}   &     ``537''           &  ``539''                 & ``543''                &                ``541''                                  \\  \hline   \hline
        System                              & ReTi$_2$$^{\star}$    & Hf$_2$Tl$^{\star}$       &  Be$_2$Zn$^{\star}$    & Re$_3$Ru$^{\star}$                                      \\ \hline
        Lattice                             & Hexagonal          &Tetragonal                & Orthorhombic           & Orthorhombic                                            \\ \hline
        Space Group                         & P$\bar{3}$m1 \#164    &I4/mmm \#139              & Fmmm  \#69             &Imm2 \#44                                                \\ \hline
        Pearson symbol                      & hP3                   &tI6                       & oF12                   &oI8                                                      \\ \hline
        Prim. vect.                         &  (SG option 2)        &   (SG option 2)          & (SG option 2)          &                                                         \\
        $a,b,c$ (\AA)                       &4.381, 4.381, 2.835    &4.422, 4.422, 7.385       &  3.780, 2.0978, 10.300 &9.005, 2.757, 4.775                                      \\
        $\alpha,\beta,\gamma$ (deg)         &90, 90, 120            &90, 72.577, 90            & 90, 90, 90             &  90, 90, 90                                             \\ \hline
        Wyckoff                             &Re1 0,0,0 (1a)         &Hf1 0,0,0.1746 (4e)       &Be1 0,0,0.17832  (8i)   & Re1 1/4,0,0 (4c)                                        \\
        positions                           &Ti1 1/3,2/3,0.299 (2d) &Tl2 0,0,1/2 (2b)          &Zn1 0,0,1/2  (4b)       & Re2 0,1/2,1/6 (2b)                                      \\
        \cite{bilbao,tables_crystallography}&                       &                          &                        & Ru1 0,0,2/3 (2a)                                        \\ \hline
        {\small AFLOW} label \cite{aflow}   &           ``545"      &    ``547''               &     ``549''            &            ``551"                                       \\   \hline   \hline
      \end{tabular}
    }
  \end{table}
\end{widetext}

\section{Conclusion}
Empirical structure maps employ correlations between
simple properties of the elements (e.g.\ position in the periodic table) and known crystal types
to predict stable ordered structures of new alloys.
In contrast, {\it ab initio} methods do not use previously available data but try to determine structure
by computing from scratch a quantum mechanical description of the system.
The structure maps presented above merge the results of the empirical Pettifor maps and
HT {\it ab initio} calculations. The HT results reveal unsuspected clusters of structures
both in binary systems known to be compound forming and in the empty gaps in the empirical maps,
that are due to lack of experimental information. It is thus demonstrated that this integration of the two
different approaches produces enhanced maps that should provide a more comprehensive improved foundation for rational materials design.

\section{Acknowledgments}
We thank Wahyu Setyawan, Mike Mehl, Leeor Kronik, and Michal Jahn\'atek for fruitful discussions.
Research supported by ONR (N00014-07-1-0878, N00014-07-1-1085, N00014-09-1-0921, N00014-10-1-0436), and NSF (DMR-0639822, DMR-0650406).
We are grateful for extensive use of the Fulton Supercomputer Center at Brigham Young University and Teragrid resources (MCA-07S005).
SC acknowledges the Feinberg support at the Weizmann Institute of Science.



\end{document}